# Ultrafast pump-probe spectroscopic signatures of superconducting and pseudogap phases in $YBa_2Cu_3O_{7-\delta}$ films

Chunfeng Zhang,[1] Wei Li,[1] B. Gray,[2] Bin He,[1] Ye Wang,[1] Fan Yang,[1] Xiaoyong Wang,[1] J. Chakhalian,[2,3,a] and Min Xiao[1,2,b]

[1]National Laboratory of Solid State Microstructures and Department of Physics, Nanjing University, Nanjing, 210093, China
[2]Department of Physics, University of Arkansas, Fayetteville, Arkansas 72701, USA
[3]Division of Physics and Applied Physics, School of Physical and Mathematical Sciences, Nanyang Technological University, Singapore 637371, Singapore

Electronic mails: [a] jchakhal@uark.edu, [b] mxiao@uark.edu

**Abstract**

Femtosecond pump-probe spectroscopy is applied to identify transient optical signatures of phase transitions in optimally doped $YBa_2Cu_3O_{7-\delta}$ films. To elucidate the dynamics of superconducting and pseudogap phases, the slow thermal component is removed from the time-domain traces of photo-induced reflectivity in a high-flux regime with low frequency pulse rate. The rescaled data exhibit distinct signatures of the phase separation with abrupt changes at the onsets of $T_{SC}$ and $T_{PG}$ in excellent agreement with transport data. Compared to the superconducting phase, the response of the pseudogap phase is characterized by the strongly reduced reflectivity change accompanied by a faster recovery time.

The interactions of fermionic quasi-particles (QPs) with bosonic excitations are fundamental for description of high temperature superconducting (SC) phase including competing electronic and magnetic phases in cuprates.[1,2] In the past, the experiments that probe the electronic properties at equilibrium have provided strong evidence for the coupling between QPs and bosonic excitations linked to lattice phonon modes and spin degrees of freedom.[1-2] Time-resolved spectroscopy with ultrafast temporal resolution is a complementary tool to investigate the temporal evolution of the QP-bosonic degrees of freedom[3-8] that is capable to disentangle the contribution of bosonic excitations from the dynamics viewpoint.[6] Based on this observation, the non-equilibrium optical spectroscopy has been extensively applied to explore mechanisms of phase separation,[9-15] boson pairing,[16] and the electron-phonon coupling[17-21] in superconductors by monitoring the QP dynamics; the experimental results are conventionally connected to microscopic processes by application of a phenomenological model proposed by Rothwarf and Taylor (*i.e.* the RT model).[15,22-23]

Conventionally, a pump-probe experiment on superconductors uses a pump pulse that breaks Cooper pairs and excites QPs to high-energy states while the second or probe pulse monitors recombination/relaxation processes of the photo-excited QPs.[24] Towards this implementation, there are a number of challenges to overcome. For example, in the previous measurements on optimally-doped (OD) YBa$_2$Cu$_3$O$_{7-\delta}$ (YBCO) samples, the observed sign of time-resolved differential reflectivity ($\Delta R/R$) was reported to be either positive or negative.[9,11,25,26] This fundamental discrepancy was attributed to the dependence of $\Delta R/R$ on several factors including laser power,[15,23] wavelength,[25,27] polarization,[25,28] and repetition rate[24,29]. Furthermore, in a recent effort to uncover the transient optical signatures of the pseudogap (PG) phase,[11-14,28,30-32] and elucidate a possible relationship between the PG state and SC state in OD YBCO [31-32], the PG phase was either identified as a weak transient feature [11,30] or found to be completely absent from the spectra.[9,19,25,26]

Here, we report on a new procedure to obtain the definitive ultrafast optical signatures of phase separation from time-resolved $\Delta R/R$ in high quality OD YBCO thin films. Temperature-dependent $\Delta R/R$ recorded under high-fluence excitation has revealed the characteristic transient features of the SC and PG phases. Specifically, the time-resolved $\Delta R/R$ at temperature below transition temperature ($T_{SC}$) exhibits an increase of reflectivity rapidly recovered within a few picoseconds entangled with a longer time reflectivity decrease caused by QP interactions with thermal excitations. After removing the slow relaxation component associated with the thermal bath, the rescaled data demonstrate abrupt changes in dynamics at specific temperatures revealing the onsets of the SC and PG phases and thus providing clear signatures of the phase transformation in ultrafast time scale. Furthermore, the new analysis yields excellent agreement with the previously reported data[15,32] based on the conventional RT model.[22]

The OD YBCO thin film sample with a thickness of 48 unit cells is grown on an atomically flat single crystal $SrTiO_3$ (001) substrate in the layer-by-layer fashion by laser MBE.[33,34] The temperature-dependent resistance of the sample is shown in Fig.1. As clearly seen, the sharp transition temperature to superconducting phase is ~ 93 K which indicates the high quality of the sample. In addition, the transition temperature ($T_{PG}$) for the pseudogap phase at around 152 K, where the resistance deviates from the linear regime, testifies for the optimal hole doping typical of high quality YBCO bulk crystals. Time-dependent reflectivity measured in the pump-probe configuration is schematically shown in the inset of Fig.1. An 1 kHz Ti:Sapphire regenerative amplifier (Libra, Coherent Inc) operating at 800 nm with a temporal duration of ~ 90 fs is used for the experiment. The pump and probe beams are configured to have cross polarizations with the signal of reflectivity change recorded by a balanced detector and analyzed by a lock-in amplifier. A liquid helium cryostat (MicroCryostatHe, Oxford) with a silica optical window is used to perform the temperature-dependent measurements.

Time-resolved $\Delta R/R$ measured in the SC phase is sensitive to the excitation power in a low-fluence regime.[15] Upon increasing the excitation value above the critical threshold $\Phi_0$, the maximum value of reflectivity change tends to saturate since no more Cooper pairs can be destroyed. To eliminate the effect due to laser power dependence, we keep the pump beam in a high-fluence regime with the excitation flux (~ 0.5 mJ/cm$^2$) set over ten times of the saturation value; In this high-fluence regime, the relaxation lifetime is no longer power-dependent.[15] In addition, we have addressed the issue of the accumulation of laser heating by operating the laser at a low repetition rate of 1 kHz. This mode of operation is in sharp contrast to the previous reports on heating issues with the laser operating at high repetition rate above 1 MHz. Under these stringent conditions, the time-resolved ΔR/R data provide a reliable dataset to investigate signatures of certain special electronic phases in cuprates.

A typical set of time-domain traces measured at different temperatures are presented in Fig. 2 (a). As one can see that, at low temperatures below $T_{SC}$, a fast increase of the reflectivity towards positive values is clearly observed, which is followed by a longer recovery crossing the zero time value. With increasing temperature above $T_{SC}$, the peak becomes markedly weaker and gradually vanishes around 160 K. The slow recovery component recorded at high temperatures above 160K is associated with the thermal equilibrium bolometric response.[9]

Previously, the QP dynamics associated with SC condensate has been a core subject for ultrafast studies on cuprates. In connection to this, the RT model was developed and widely accepted to interpret the transient optical data in the SC phase. It is interesting to note that even though the RT model was originally devised to incorporate the phonon-mediated QP recombination term, it can be further extended to include the spin-fluctuation mediated recombination if the phonon term is regarded as a general boson mode. On the other hand, since the RT model only concerns the recombination and annihilation of Cooper pairs, the entanglement between QP dynamics for broken Cooper pairs and bolometric response due to thermal equilibrium in the traces makes direct application of the model unsuitable here [see Fig. 2(a)]. To address this issue, we have reanalyzed the data by removing the contribution due to thermal excitations. More specifically, first we analyze the traces at high temperatures (>170 K) with a phenomenological multi-exponential function consisting of a bi-exponential decay component and an exponential recovery component in the form

$$G(t) = C * g(t) = C(-e^{t/\tau_0} - g_1 e^{t/\tau_1} + g_2 e^{t/\tau_2}). \qquad (1)$$

As seen in Fig. 2 (a) (the black curves), all the high temperature data above 160 K can be well reproduced by $g(t)$ with fixed parameters of $\tau_0 \sim 1.4 ps$, $\tau_1 \sim 10.8 ps$, $\tau_2 \sim 126.2\,ps$, $g_1 \sim 1.12$, and $g_2 \sim 1.73$ and a single fitting parameter. We can then exclude the component of thermal equilibrium to extract the QP dynamics at the low temperature. In turn, the parameter $C$ is determined by fitting the time-dependent traces at long time (e.g., t > 30 ps) with the function $G(t)$ since the dynamical feature for the annihilation and recombination of Cooper pairs is in a shorter temporal scale.

To test the validity of the new procedure, we apply the conventional RT model to simulate the rescaled data. In the high-fluence regime, the model can be described by the set of the time-dependent RT equations (RTEs) as:[15,22]

$$dn(t)/dt = I_{QP}(t) + 2\gamma P(t) - \beta n^2(t)$$
$$dP(t)/dt = I_{ph}(t) - \gamma P(t) + \frac{\beta}{2}n^2(t) - \gamma_{esc}(t)(P(t) - P_T) \quad (2)$$

the density of photo-excited QPs ($n(t)$) coupled to boson modes (*e.g.* phonons) and, $P(t)$ being the density of gap-energy phonons. The photo-injected QPs and phonons are represented by the terms of $I_{QP}(t)$ and $I_{ph}(t)$ with Gaussian temporal profiles, respectively. The processes of Cooper pair annihilation due to phonon absorption and recombination of QPs to form Cooper pairs with phonon emission are given by the terms of $\gamma P(t)$ and $\beta n^2(t)$, respectively. The escape of gap-energy phonons is the term of $\gamma_{esc}(t)(P(t) - P_T)$, with $P_T$ being the thermal phonon density. $\Delta R/R$ is assumed to be proportional to the solution of $n(t)$. As clearly seen in Fig. 2(b), the rescaled data are remarkably well reproduced by the RTEs for a *d*-wave gap symmetry with $\gamma_{esc}(t) = \gamma_{esc}(0)[1 - an(t)^{3/2}]^{2\alpha}$. The obtained best fitted parameters ($\gamma = 4.5\,ps^{-1}$, $\beta = 0.1\,cm^2/s$, $\alpha = 1$, and $a = 1 \times 10^{-20}\,cm^{-3}$) are also in good agreement with the previously reported values,[15,23,27] thus indicating that our scaling procedure is well suited for extracting the QP dynamics associated with Cooper pairs.

With the above approach in hand, we then analyze the temperature dependence of the rescaled data to investigate the presence of optical signatures for the SC and PG phases in OD YBCO [see Fig. 3(a)]. Remarkably, all the treated traces fall into three distinct regimes: (*i*) when T > 160 K, the thermal equilibrium strongly dominates the transient optical response, (*ii*) below $T_{SC}$, the data shows a very rapid positive reflectivity change followed by a slow exponential recovery and (*iii*) in the intermediate temperature regime of $T_{SC} < T < T_{PG}$, the negative reflectivity change is progressively weakened. These distinct features can be associated with the known phase separation regimes of YBCO. To quantify the abrupt changes, in Fig. 3(b) we plot the maximum ΔR/R values as a function of temperature. By comparing to the transport data shown in Fig 1, in traversing T = 93 K, we observe a four-fold increase in ΔR/R which is a clear optical signature of developing the SC phase transition. The second jump occurs at around ~ 160 K, which is remarkably close to the onset of $T_{PG}$ derived from the transport measurement (152 K). In comparison to the SC phase, this regime is characterized by a weaker reflectivity change and faster recovery time. Based on these observations, the second change can be assigned to the PG phase.

For almost three decades, YBCO has been a model high $T_{SC}$ superconductor for ultrafast optical studies, where the PG phase is clearly observed in under-doped samples.[31-32] However, for OD YBCO samples, which is the subject of this study, the PG feature was either unobserved[9,19,25,26] or, in a very few cases,[11,30] reported as a weak transient clue of the pseudogap. To explain those earlier results, we note that, apart from the quality of the samples, the relatively weak laser excitation could be another important factor. Specifically, with an increase in the excitation flux, the SC condensate response increases more rapidly and saturates at substantially lower power than in the PG phase.[13] In previous studies of OD cuprates, the excitation pulse energy was kept at very low levels. Under such weak excitation, the response of PG is not clearly resolvable since it is strongly convoluted with the bolometric reflectivity change showing in the opposite direction. In contrast, the intense pulses combined with the low repetition rate used in this study are clearly superior for revealing the strong response in the pseudo-gap regime. While the origin of such PG phase in cuprates is still under active debate,[35-37] recently the incoherent pair formation has been identified in PG phase.[38] By virtue of the same observation, we conjecture that perhaps similar to the saturated response of the SC phase, the pulse energy in a high-flux regime ($> 10\,\Phi_0$) seems sufficient to break the majority of incoherent pairs in the PG phase.

In summary, we have performed an ultrafast optical study on the OD YBCO thin films with excitation power in the high-flux regime. We observed unambiguous ultrafast optical signatures of the PG and SC phases in the rescaled traces with eliminated bolometric response. In comparison to the SC phase, PG is characterized by less pronounced change in reflectivity but with a much faster recovery time. The obtained optical signatures are likely generic for the high $T_{SC}$ cuprate family. The new method should be useful for understanding the enigmatic problem of high $T_{SC}$ superconductivity and the role of the PG phase from the dynamics point of view.

The work at NJU is supported by the National Basic Research Program of China (2012CB921801 and 2013CB932903, MOST), the National Science Foundation of China (9123310, 61108001, 11227406, and 11021403), and the Program of International S&T Cooperation (2011DFA01400, MOST). J.C. is supported by NSF Grant No. DMR-0747808 and partially by DOD-ARO Grant No. 0402-17291. C.Z. is supported by the New Century Excellent Talents program (NCET-09-0467), Fundamental Research Funds for the Central Universities, and the Priority Academic Program Development of Jiangsu Higher Education Institutions (PAPD). C.Z. acknowledges Prof. Jianxin Li for stimulated discussions. J.C. deeply acknowledges discussions with Prof. Richard Averitt.

**Figure Captions**

**Figure 1**, Temperature-dependent resistance of the YBCO film. The transition temperatures of the superconducting phase and the pseudogap phase are about 93 K and 152 K, respectively. Inset shows a schematic of the time-resolved pump-probe experiments.

**Figure 2**, (a) The measured signal $\Delta R/R$ of as a function of delay time at different temperatures. The solid black lines are the fitted curves including the thermal equilibrium function. The pump flux is $\sim 0.5$ mJ/cm$^2$. (b) An example of the treated data after subtraction of the thermal equilibrium contribution. The rescaled data can be well reproduced by the conventional RT model.

**Figure 3**, (a) The rescaled $\Delta R/R$ data at different temperatures. Abrupt changes occur at the temperatures around $T_{SC}$ and $T_{PG}$. (b) The maximum change at zero delay is plotted as a function of temperature.

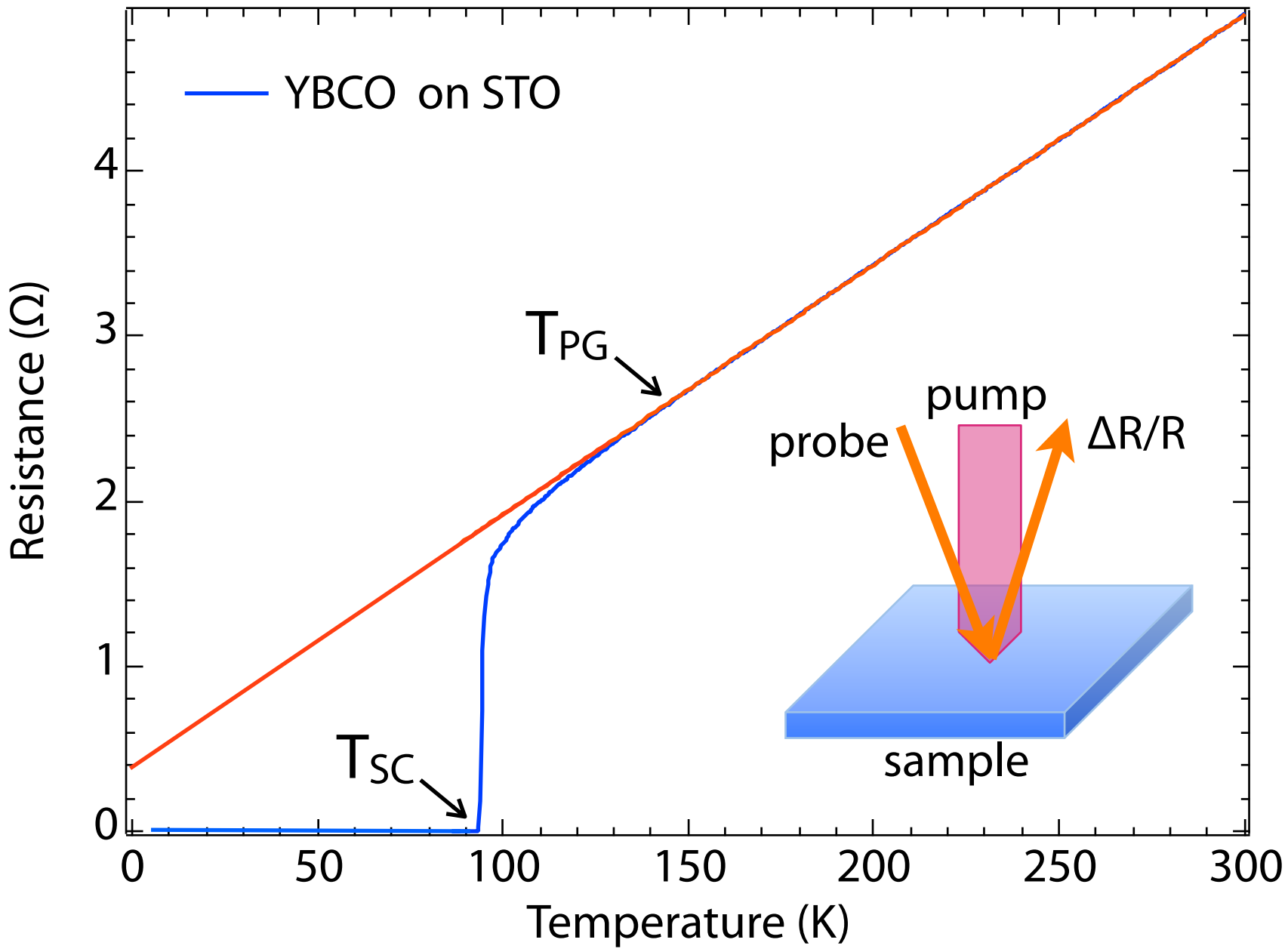

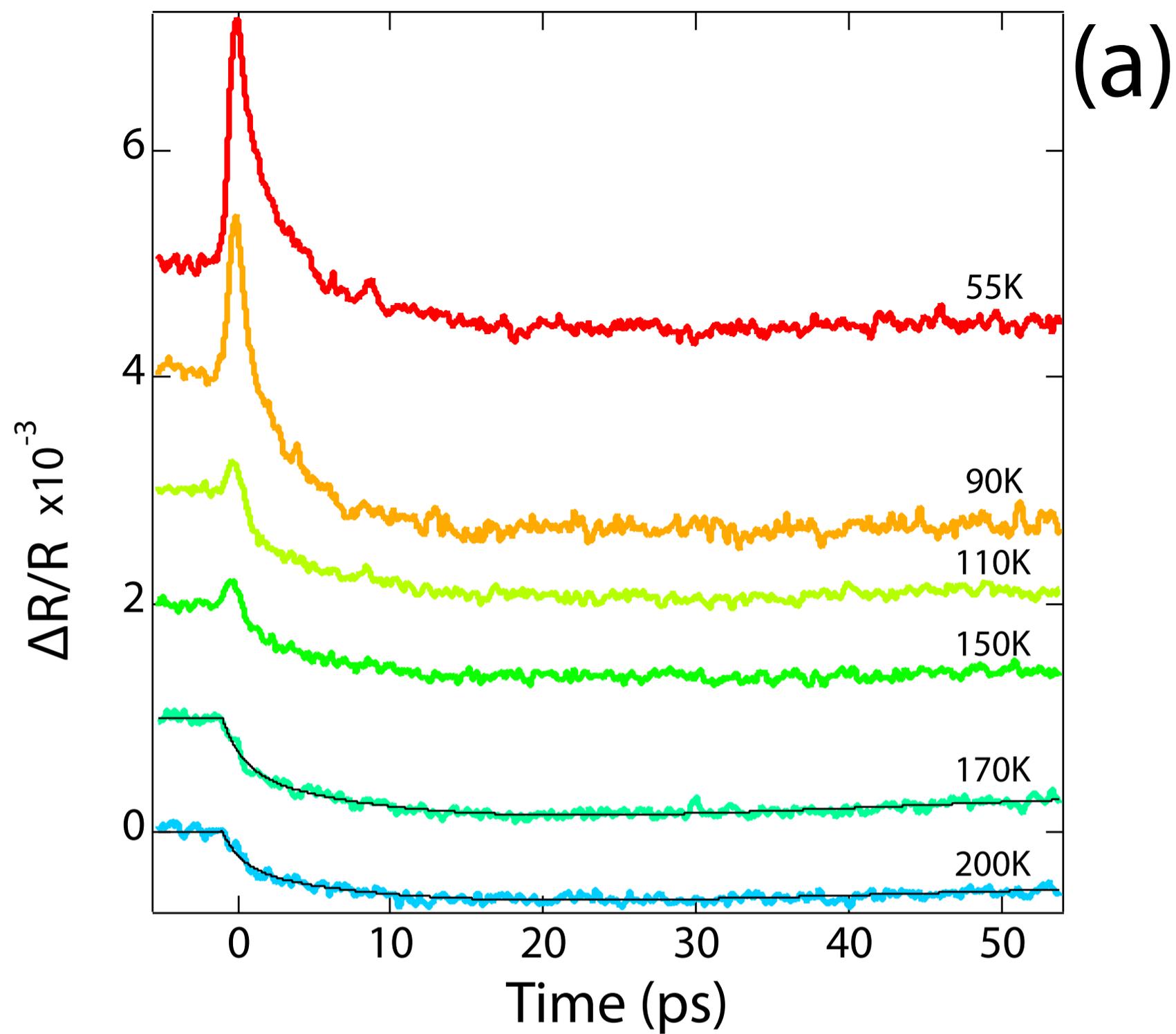

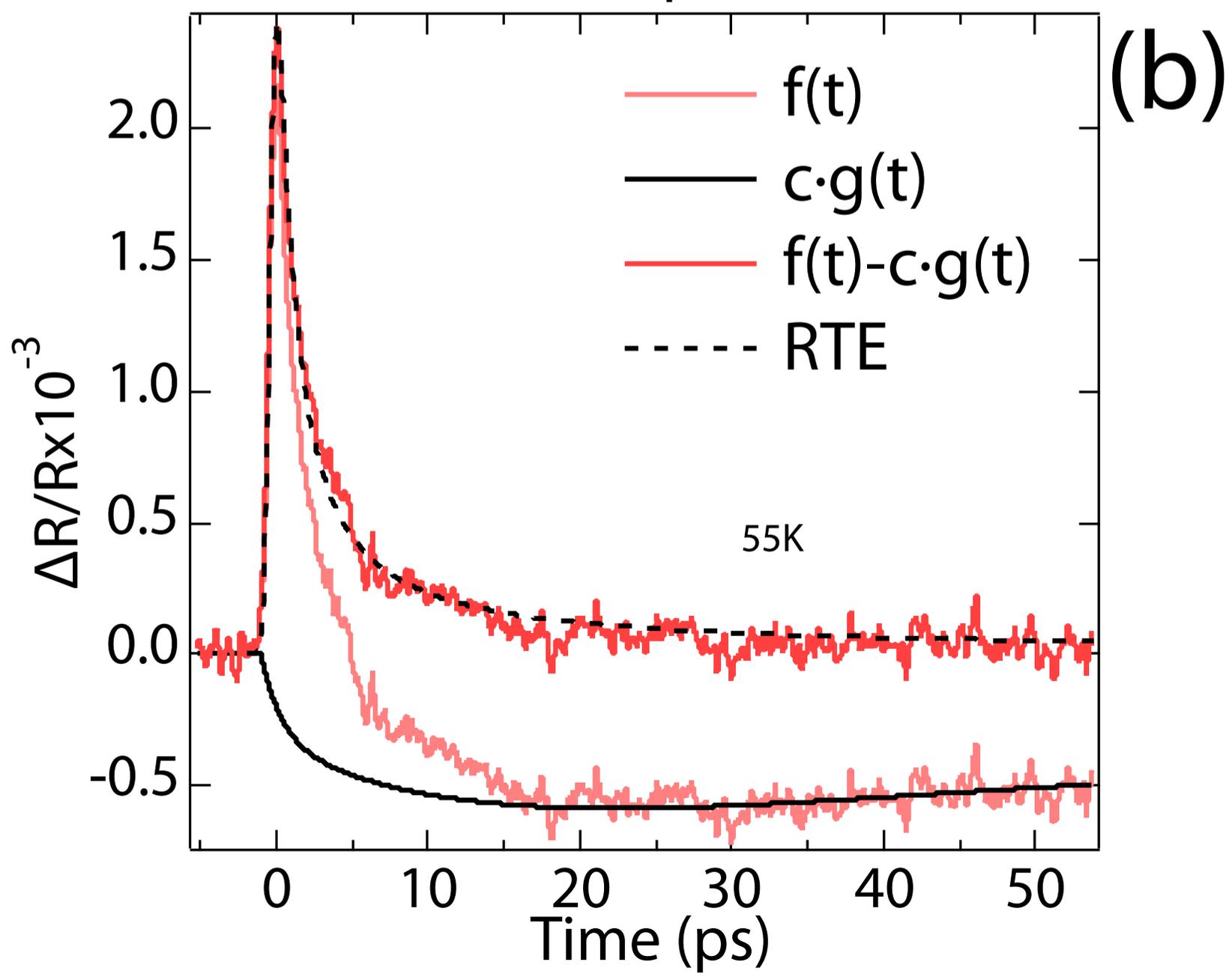

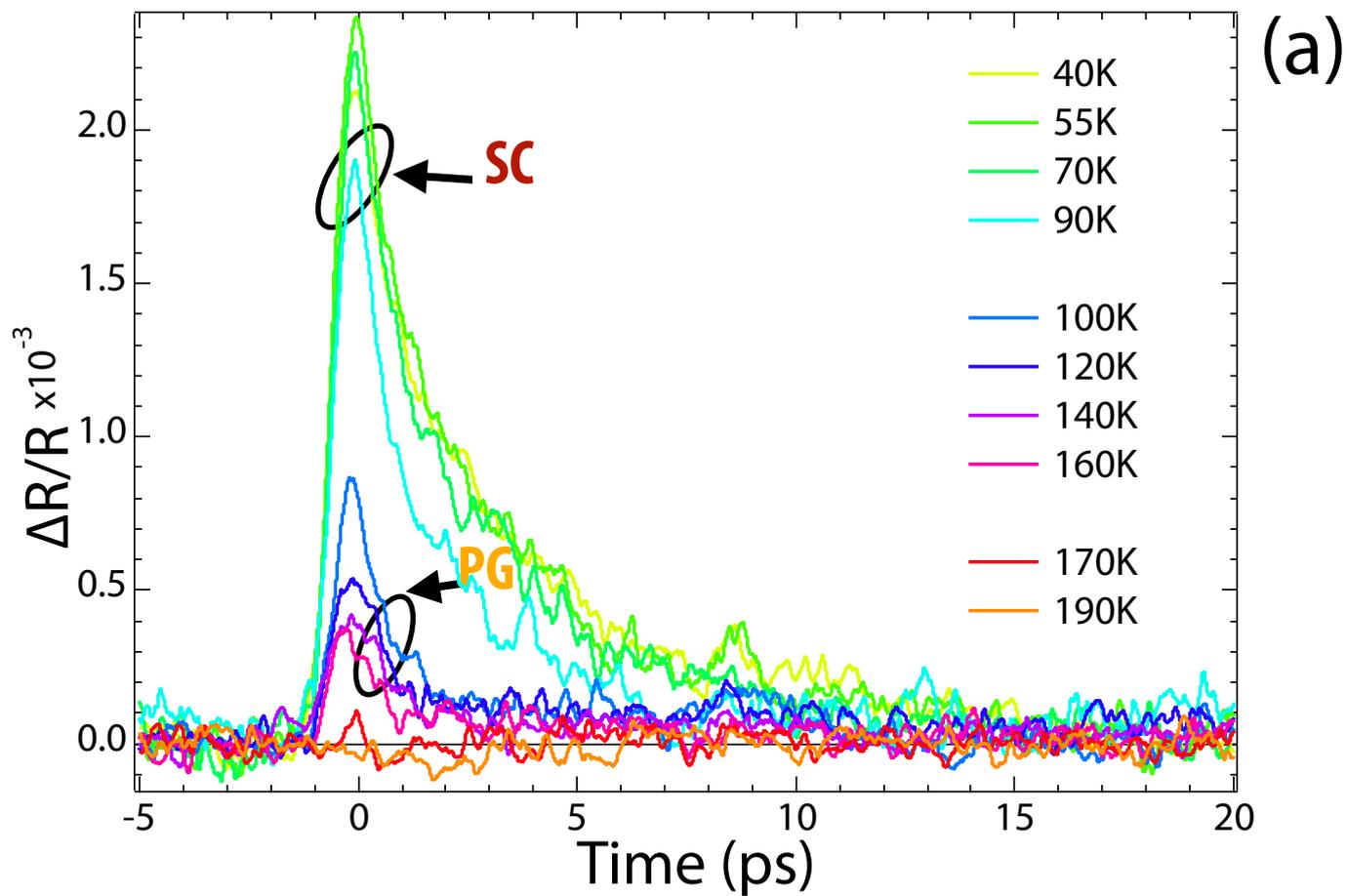

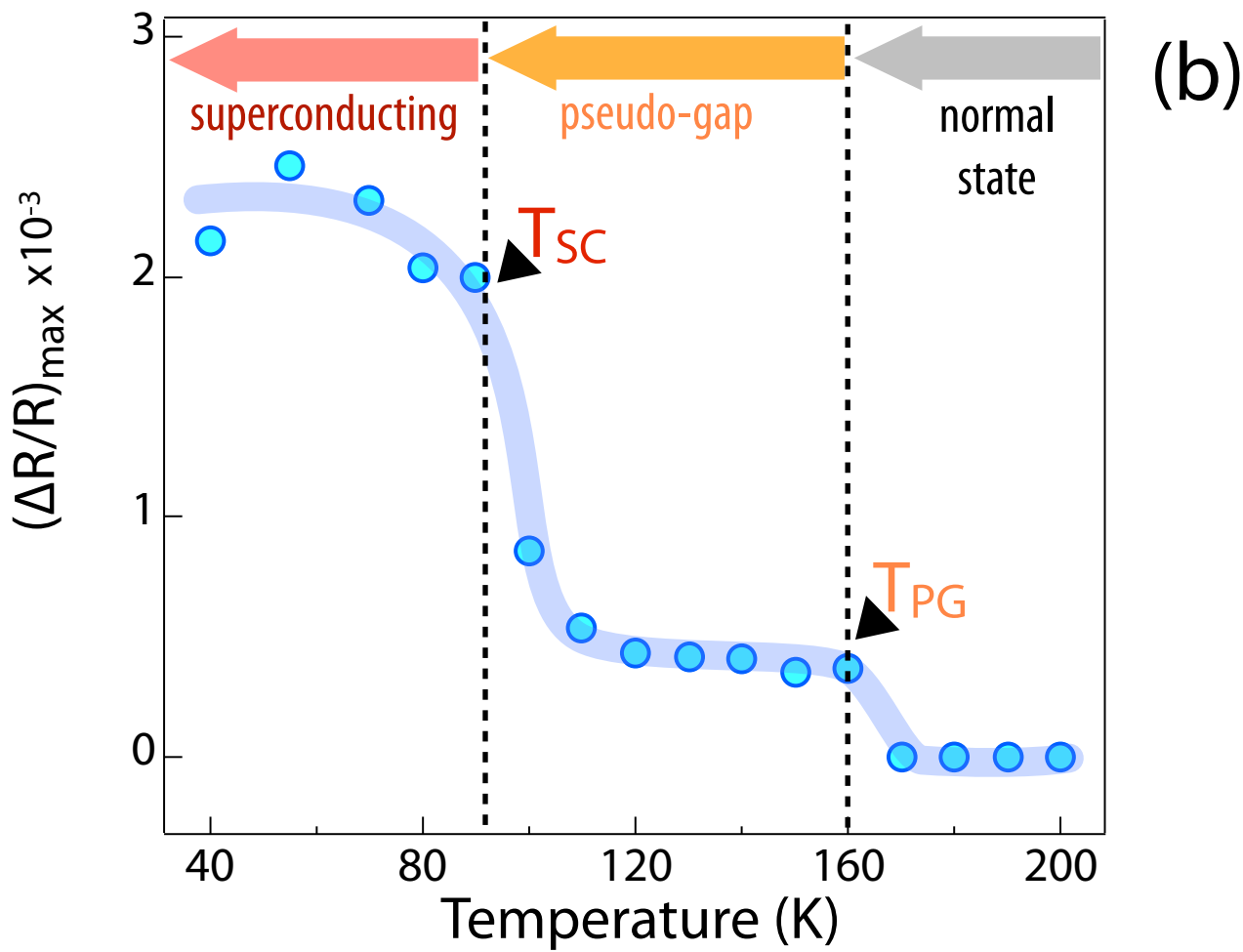